\pdfoutput=1
\def\year{2021}\relax
\documentclass[letterpaper]{article} % DO NOT CHANGE THIS
\usepackage{aaai21}  % DO NOT CHANGE THIS
\usepackage{times}  % DO NOT CHANGE THIS
\usepackage{helvet} % DO NOT CHANGE THIS
\usepackage{courier}  % DO NOT CHANGE THIS
\usepackage[hyphens]{url}  % DO NOT CHANGE THIS
\usepackage{graphicx} % DO NOT CHANGE THIS
\usepackage{natbib}  % DO NOT CHANGE THIS AND DO NOT ADD ANY OPTIONS TO IT
\usepackage{caption} % DO NOT CHANGE THIS AND DO NOT ADD ANY OPTIONS TO IT
\usepackage{subfigure}
\usepackage{multirow}
\usepackage{mathtools}
\usepackage{enumitem}
\usepackage{amsmath}
\title{Partisan Asymmetries in Exposure to Misinformation}
\author{
    Ashwin Rao, Fred Morstatter, Kristina Lerman \\
    Information Sciences Institute\\University of Southern California
}
\affiliations{
}

\begin{document}

\maketitle

\begin{abstract}
Health misinformation is believed to have contributed to vaccine hesitancy during the Covid-19 pandemic, highlighting concerns about the role of social media in polarization and social stability. While previous research has identified a link between political partisanship and misinformation sharing online, the inter- action between partisanship and how much misinformation people see within their social networks has not been well studied. As a result, we do not know whether partisanship drives exposure to misinformation or people selectively share misinformation despite being exposed to factual content. We study Twitter discussions about the Covid-19 pandemic, classifying users ideologically along political and factual dimen- sions. We find partisan asymmetries in both sharing behaviors and exposure, with conservatives more likely to see and share misinformation and moderate liberals seeing the most factual content. We identify multi-dimensional echo chambers that expose users to ideologically congruent content; however, the in- teraction between political and factual dimensions creates conditions for the highly polarized users---hard- line conservatives and liberals---to amplify misinformation. Despite this, misinformation receives less at- tention than factual content and political moderates, who represent the bulk of users in our sample, help filter out misinformation, reducing the amount of low-factuality content in the information ecosystem. Identifying the extent of polarization and how political ideology can exacerbate misinformation can po- tentially help public health experts and policy makers improve their messaging to promote consensus. 

%Social media has emerged as a catalyst of political polarization. As the COVID-19 pandemic continues to wreak havoc around the world, healthcare experts and policy makers grapple to build consensus between polarized factions. The proliferation of COVID-19 misinformation on social media platforms has arrested the efficacy of mitigation efforts with a significant fraction of the society choosing to disobey masking mandates and disregard calls for inoculation. Political conservatism has been shown to be correlated with such anti-science attitudes. Analyzing COVID-19 discourse from 2.4M users on Twitter, we propose a method to leverage users' domain sharing behaviors to quantify their propensity to misinformation and find that political conservatism is correlated with propensity to misinformation. We find that users are surrounded by like minded peers who share not just their political beliefs but also opinions towards content factuality, indicating the presence of echo chambers. We also find that at political extremes, users generate lower factual content than what they are exposed to thereby, having a higher propensity to misinformation. Additionally, we find that the intensity of belief in a political ideology better explains propensity to misinformation for conservatives as opposed to liberals. These findings throw light on the multidimensionality of echo chambers on social media and more specifically, help identify sources of misinformation and factors that facilitate it. 

\end{abstract}

%\noindent As Covid-19 pandemic continues to spread around the world, misinformation about the pandemic---its toll, severity, efficacy of mitigation measures like masks and vaccines---abounds online.  

\section{Introduction}

The growing popularity of social media as a source of news for a large portion of the population~\cite{pew2018social_news} has raised concerns about the quality and validity of information being shared online and its effect on polarization, which refers to division of the public  into two groups with sharply contrasting opinions or beliefs~\cite{levy2021social,van2021social}.  These concerns have only grown in urgency with the emerging evidence that social media enables the spread of misinformation and politically polarized content about the Covid-19 pandemic, its toll, mitigation measures, and the efficacy of interventions, therapies and vaccines~\cite{jiang2021social,rao2021political}. According to a Pew Report~\cite{pew2020partisan}, political ideology explains a partisan divide in attitudes about Covid-19 and compliance with health guidelines~\cite{gollwitzer2020partisan}. Since effective response to the pandemic requires collective action, e.g., mass vaccination to achieve herd immunity, social media can exacerbate public health impacts of the pandemic by deepening societal divisions and amplifying health misinformation~\cite{roozenbeek2020misinfo,chen2021covid, memon2020characterizing}, thereby hindering consensus.

Multiple studies have examined how misinformation and ``fake news'' are shared online (e.g., see \citet{Grinberg374,vousoghi}), focusing on methods to automatically recognize misinformation~\cite{pennycook2019fighting} and characterize people who spread it~\cite{Pennycook2521}. 
Online polarization has been a research topic for an even longer time period. Studies have shown that people share information that aligns with their attitudes and political beliefs~\cite{levy2021social}. These attitudes can be measured by analyzing online activity traces, based on the text of the messages people share~\cite{Conover2012} or the links to news sources embedded in their posts~\cite{rao2021political}. 
People also seek out information sources that are consistent with their beliefs~\cite{knobloch2009looking}, following and retweeting social media  partisans who have similar ideology to their own~\cite{Barbera2015,badawy2018analyzing}. %Selective exposure to ideologically congruent online content  
These activities facilitate the development of ``echo chambers,'' which surround people with like-minded peers who confirm their pre-existing attitudes and beliefs. Studies have explored political echo chambers and measured their effects~\cite{Cinellie2023301118,bakshy2015exposure,nikolovright,jiang2021social}, but their role in exposure misinformation, especially in the context of the pandemic, has not yet been explored. %. Echo chambers are considered to be detrimental as they amplify polarization, and in the case of COVID-19, reinforce maladaptive behaviors. 

% interactions
Previous works have identified a link between partisanship and misinformation: politically conservative social media users are more likely to spread misinformation~\cite{Grinberg374,nikolovright} and anti-science content~\cite{rao2021political}. This link partly explains the opposition by conservatives to Covid-19 mitigation measures~\cite{gollwitzer2020partisan}. 
However, the interaction between partisanship and exposure to misinformation has not been as well studied. As a result, we do not know whether partisanship drives selective exposure to misinformation or people selectively share misinformation despite being exposed to  diverse and high quality information sources.

In the context of a global pandemic and the public's increasing reliance on online information, it is important to understand the factors shaping public's exposure to polarized information and misinformation. We organize our research around these questions:

\begin{description}
    \item[RQ1] Is polarization of information people share (along partisan and factual dimensions) aligned with the polarization of information they see within their social networks? Are there multidimensional echo chambers?
    \item[RQ2] How well are the dimensions of polarization correlated, i.e., how much is partisanship correlated with content factuality?
    \item[RQ3] Are exposures to misinformation asymmetrical along partisan lines?
    \item[RQ4] Are there partisan asymmetries in the selective amplification or filtering of misinformation?
    \item[RQ5] Do people pay more attention to factual content or misinformation?
\end{description}

Our study addresses this challenge by examining online discussions about the Covid-19 pandemic. First, we classify social media users ideologically along political and factual dimensions, assigning them a multi- dimensional polarization score. Next, we quantify multi-dimensional polarization of the information users see in their friends’ posts. As a proxy of friends, we take the messages posted by accounts the user retweets. We identify echo chambers that expose users to ideologically congruent information along political and factual dimensions. While social media users tend to surround themselves with peers who share similar views on politics and factuality, there are partisan asymmetries in exposure to factual content. Additionally, the substantial interaction between the two dimensions, also observed in earlier studies \cite{Grinberg374}, creates conditions for politically polarized users to amplify misinformation. These polarized users, who represent hardline partisans on both sides of the political spectrum, selectively share misinformation. However, such users receive less attention than those sharing factual content, and political moderates, who represent the bulk of users in our study, help filter out misinformation, reducing the amount of low-factuality content in the information ecosystem. Identifying the extent of polarization and how political ideology can exacerbate misinformation can potentially help public health experts and policy makers improve their messaging to facilitate consensus and compliance with public health measures.

%Identifying the extent of polarization and how political ideology can exacerbate misinformation can potentially help public health experts and policy makers improve their messaging to facilitate consensus and compliance with public health measures. 

%we also find that political conservatism is correlated with propensity to misinformation. However, we  identify the presence of an off-diagonal comprising of hardline liberals who have a predilection for misinformation and a fraction of moderate conservatives who generate highly factual content. Despite the presence of this off-diagonal, we see that political ideology better explains the variance in misinformation susceptibility for conservatives as compared to liberal leaning users. Identifying the extent of polarization and how political ideology can exacerbate misinformation can potentially help public health experts and policy makers send out targeted messaging to bring about consensus and awareness. 

\section{Related Work}
Researchers define polarization as divergence of opinions along the political dimension and study its impact on other opinions, such as scientific topics~\cite{bessi2016users}. However, opinions on controversial topics are often correlated \cite{baumann2020modeling}. For example, those who oppose lockdowns as a way to suppress the spread of the disease also resist vaccinations. To capture some of the complexity of polarization we project opinions in a multi-dimensional space, with different axes corresponding to different semantic dimensions. Once we identify the dimensions of polarization and measure them, we can study dynamics of polarized opinions, their interactions and regional differences.

Label propagation leverages structure of connections in a network to infer the political views based on the ideology of the accounts users  retweet (cf. \cite{badawy2018analyzing}). The intuition behind the approach is that people prefer to connect to (e.g., retweet content posted by) others who share their opinions and ideology~\cite{boyd2010tweet,metaxas2015retweets}. 

Others have looked at echo chambers in the context of online discourse. The authors in ~\cite{Cinellie2023301118} studied the effect of echo chambers across different platforms, finding that Facebook is more segregated than other platforms. Interestingly, they find that those platforms that allow users to adjust their feed (e.g., Reddit) afford more balanced media consumption than those that don't (e.g., Twitter and Facebook). Our work builds on this by providing a multidimensional understanding of the exposure of particular users to certain content. In another work, ~\cite{nikolovright} finds that misinformation sharing is strongly correlated with right leaning by studying the network and content. This finding is echoed in~\cite{rao2021political}. This is an important departure from our work. We find that left leaning folks also produce content with low factuality. ~\cite{rao2021political} also finds that the number of moderate, factual users greatly surpasses the number of misinformants. We see this in our data as well, with more factual users as demonstrated in Fig.~\ref{fig:factlean}.

\section{Methods}

\subsection{Data}

In this study, we use the publicly available dataset \cite{chen2020tracking} comprising of 260.6M tweets related to Covid-19 posted between January 21 and July 31, 2020. These tweets contain at least one of a predetermined set of Covid-19-related keywords (e.g., coronavirus, pandemic, Wuhan, etc.). However, less than 1\% of the tweets have geographic coordinates associated with them. We therefore rely on the geolocation method\footnote{\url{https://github.com/julie-jiang/twitter-locations-us-state}} employed in \cite{jiang2020political} to determine if the user is within the US. The method works by first extracting the mentions of city or state users frequently have in their profile before employing a fuzzy matching algorithm to match them to their respective states in the US. %In order to perform this mapping, the method maintains a mapping of popular cities and state abbreviations to their state names. 
A manual review of this approach found it to be effective in identifying user's home state. This leaves us with 48M tweets generated by 2.4M geolocated users in the United States.

\subsection{Dimensions of Polarization}

In this study, we characterize attitudes along two dimensions: \textit{political} and \textit{factual}. The political dimension captures user political ideology or partisanship, ranging from hardline liberal to hardline conservative, while the factual dimension quantifies user's predilection for factual content or  misinformation. With Media Bias-Fact Check\footnote{\url{http://mediabiasfactcheck.com}} providing an exhaustive list of domains and their ideological polarities, previous studies have leveraged users' domain sharing behaviors on Twitter \cite{Cinellie2023301118,le2019measuring, rao2021political} to quantify ideological alignment.  Along the political scale, Media Bias-Fact Check lists over 2K pay-level domains (PLDs) under five mutually exclusive categories: \textit{Left}, \textit{Center-Left}, \textit{Least-Biased/Center}, \textit{Center-Right} and \textit{Right}. In addition, it also provides a measure of reporting quality for over 3.5K pay-level domains belonging to one of the six content factuality classes: \textit{Very Low}, \textit{Low}, \textit{Mixed}, \textit{Mostly Factual}, \textit{High} and \textit{Very High}. Sources generating pro-science content are categorized as \textit{High} or \textit{Very High} while, sources propagating misinformative, questionable content or anti-science content are categorized as \textit{Low} or \textit{Very Low} on the factuality scale. On the other hand, highly partisan news sources such as \textit{foxnews.com, cnn.com, huffpost.com} generally have a chequered quality of reporting and have been listed as \textit{Mixed}. Table \ref{tab:Domain_Score} refers to the collection of information sources and their ideological biases.

\begin{table}[tbh]
\centering
    \footnotesize
\begin{tabular}{p{1.2cm} p{2.4cm} p{3.75cm}}\hline
  \textbf{Dimension} & \textbf{Polarity} & \textbf{Pay-Level Domains} \\
  \hline
  \multirow{10}{*}{Politics} & \multirow{2}{*}{Left $(0)$} & 
  cnn.com, huffpost.com, dailybeast.com,$\ldots$ (350+ PLD s)\\ 
  & \multirow{2}{*}{Center-Left $(0.25)$} & aljazeera.com, independent.co.uk, lincolnproject.us $\ldots$ (500+ PLDs)\\
  & \multirow{2}{*}{Center $(0.5)$} & gallup.com, pewresearch.co.uk, wikipedia.com $\ldots$ (500+ PLDs)\\
  & \multirow{2}{*}{Center-Right $(0.75)$} & bostonherald.com, chicagotribune.com, wsj.com $\ldots$ (250+ PLDs)\\
  & \multirow{2}{*}{Right $(1)$} & foxnews.com, gppusa.com, thenationalherald.com $\ldots$ (250+ PLDs)\\
  \hline
    \multirow{15}{*}{Factuality} & \multirow{2}{*}{Very Low $(0)$} & counterthink.com, biggovernment.news, vaccines.news $\ldots$ (180+ PLDs)\\ 
    &  \multirow{2}{*}{Low $(0.2)$} & 911truth.org, althealth-works\-.com, naturalcures.com $\ldots$ (600+ PLDs)\\
    &  \multirow{2}{*}{Mixed $(0.4)$} & breitbart.com, buzzfeed.com, independent.co.uk $\ldots$ (1000+ PLDs)\\
    &  \multirow{2}{*}{Mostly Factual $(0.6)$} & drudgereport.com, washingtonpost.com, bloomberg.com $\ldots$ (200+ PLDs)\\
    &  \multirow{2}{*}{High $(0.8)$} & azcentral.com, bbc.com, nbcnews.com $\ldots$ (1300+ PLDs)\\
    &  \multirow{2}{*}{Very High $(1)$} & nationalacademyofsciences.org, nature.com, bmj.com $\ldots$ (200+ PLDs)\\
  \hline
\end{tabular}
\caption{Curated pay-level domains and their polarity scores along political and factual dimensions. For the political dimension, $\{0,0.25,0.5,0.75,1\}$ represents \textit{Left}, \textit{Center Left}, \textit{Center/Unbiased}, \textit{Center Right} and \textit{Right} sources respectively. Along the factuality or misinformation dimension, \textit{Very Low}, \textit{Low}, \textit{Mixed}, \textit{Mostly Factual}, \textit{High} and \textit{Very High} are quantified as $\{0,0.2,0.4,0.6,0.8,1\}$ in the same order}.
\label{tab:Domain_Score}
\end{table}

%In order to quantify user leaning and exposure along the two dimensions of interest we follow a two step process:

% \begin{enumerate}[label=(\alph*)]
% \item We extract tweets and retweets containing URLs. We then make use of \textit{tldextract} \footnote{\url{https://pypi.org/project/tldextract/}} to extract pay-level domains from URLs. From this subset, we filter out tweets and retweets containing pay-level domains that are not categorized under either of the two ideological polarities of interest \ref{tab:Domain_Score}. This filtered subset is then grouped by users to aggregate all PLDs generated by each user and represent their \textbf{\textit{ideological leaning}}.

% \item From the original dataset, we build a mapping of users and the accounts they retweet. This relationship provides us with a retweet network and thereby, a retweet neighborhood for each user. Previous studies have shown that users tend to retweet content that matches their ideological leaning ~\cite{boyd2010tweet,metaxas2015retweets}. Drawing from conclusions made in these studies, we argue that all content generated by users in the retweet neighborhood of a user is potentially visible to the user and thereby, contributing to the influence on the user. We extract tweets and retweets generated by users in the retweet neighborhood of each user. We then extract PLDs as done in (a) and filter out ones that do not have a listed political or factual polarity. Each user, now has a list of PLDs which quantify the \textbf{\textit{ideological exposure}} they are under. 

% \end{enumerate}

\subsection{Measuring Polarization}
We measure ideological polarization by looking at the political and factual scores of domains people share in their posts and see their friends share.

%\subsubsection{Quantifying individual leanings}
\subsubsection{Individual Polarization (Information Sharing)}
We extract tweets containing URLs and use \textit{tldextract} \footnote{\url{https://pypi.org/project/tldextract/}} to extract pay-level domains from them. We filter out tweets and retweets containing pay-level domains that are not categorized under either of the two ideological polarities of interest (Table~\ref{tab:Domain_Score}). %This filtered subset is then grouped by users to aggregate all PLDs generated by each user and represent their \textbf{\textit{ideological leaning}}.

\begin{figure}[tb]
    \centering
    \subfigure[Political Leaning Score]
    {\includegraphics[width=0.9\linewidth]{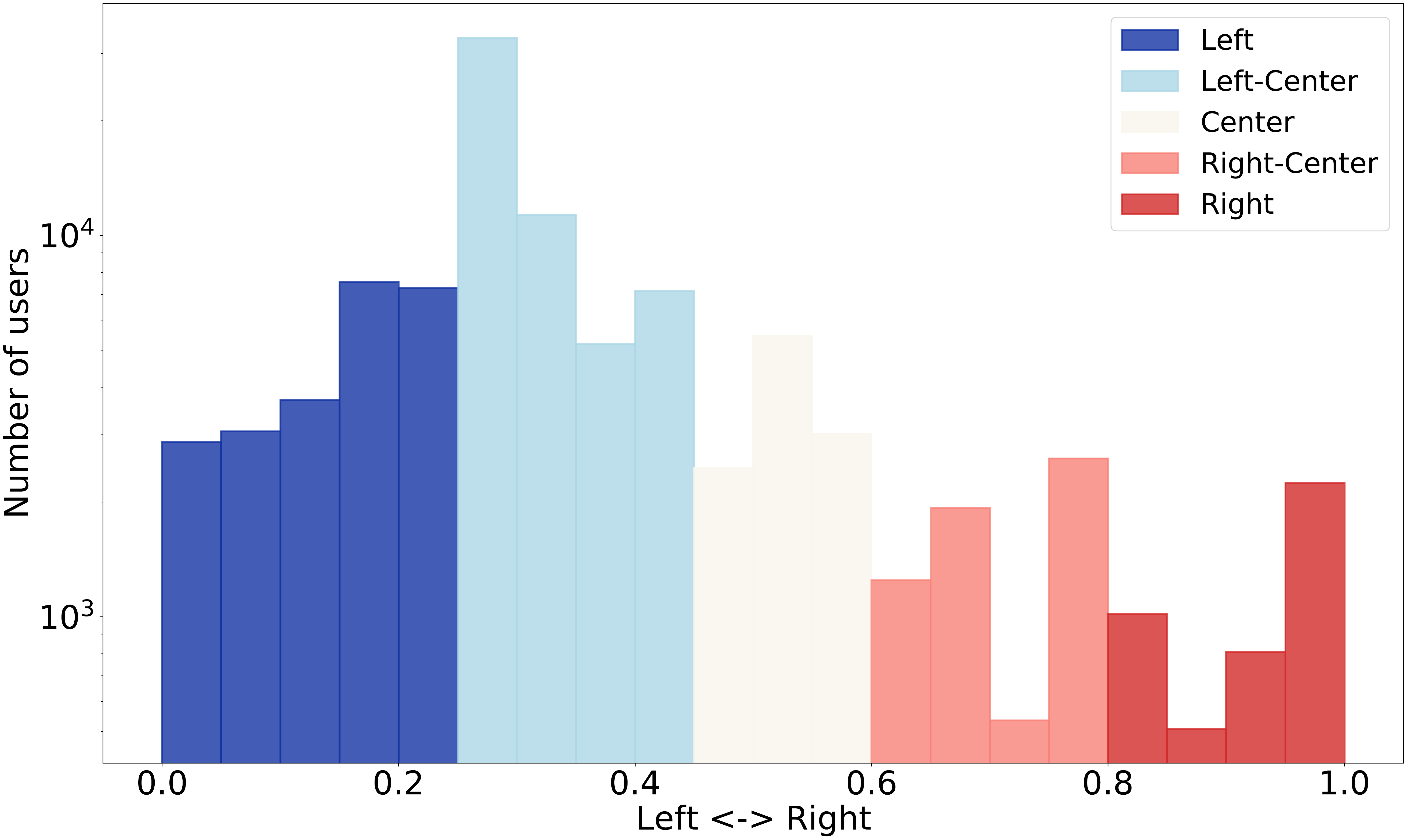}}
    \subfigure[Factual Leaning Score]{\includegraphics[width=0.9\linewidth]{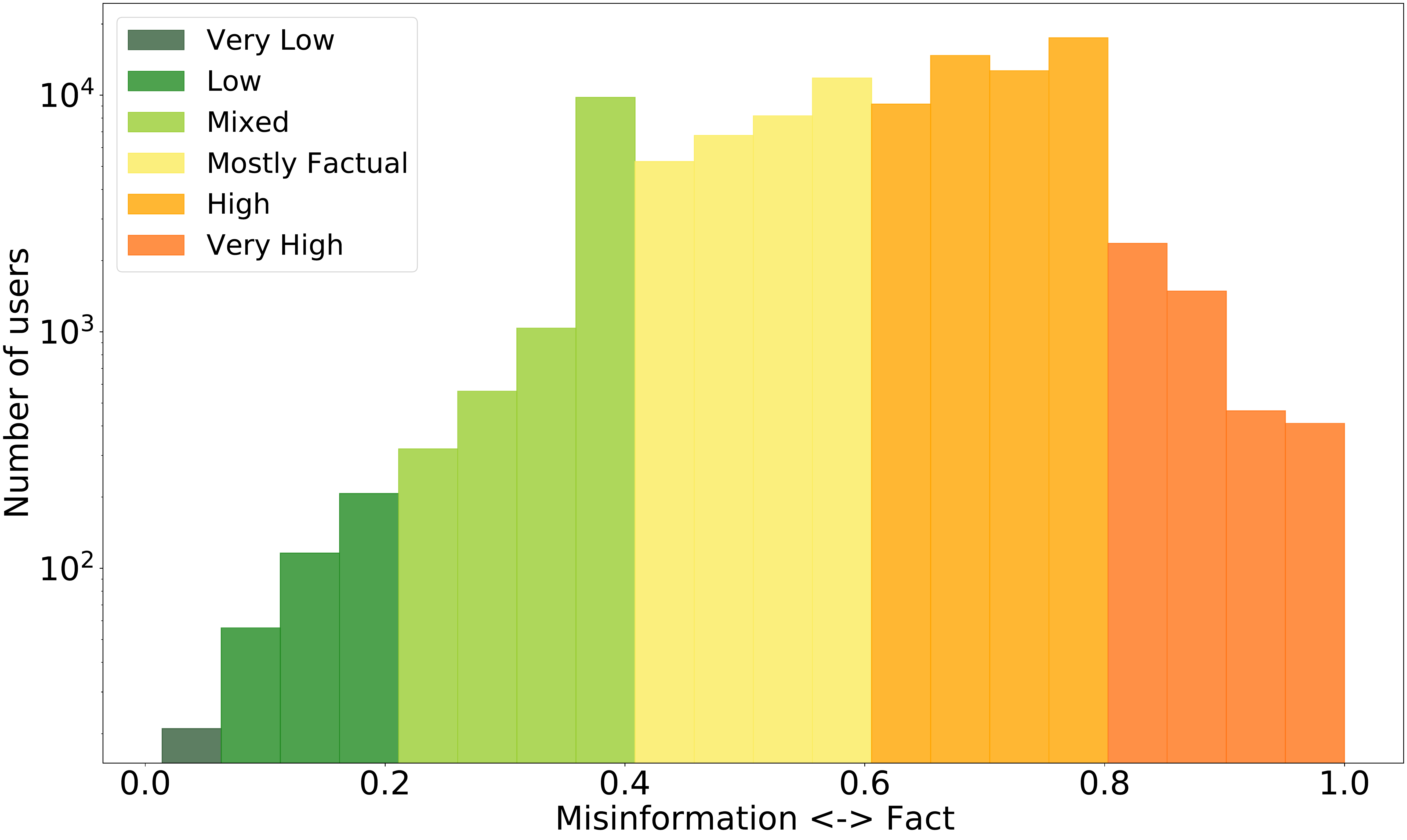}
    \label{fig:factlean}
    }
    \caption{(a) Distribution of Political Leaning domain scores. (b) Distribution of Factual Leaning domain scores.}
    \label{fig:domain-scores}
\end{figure}

Similar to previous works~\cite{Cinellie2023301118,rao2021political}, we infer a user's partisanship by averaging over the political scores of the PLDs the user shared. Likewise, we infer individual's factual scores by averaging the factual scores of the PLDs the user shared. This makes our measure of factual sharing similar to the propensity, or vulnerability, to misinformation used in previous works~\cite{Grinberg374,nikolovright}. It is important to note that individual scores quantify the information that users \textit{generate} within the online information ecosystem; therefore, users with low factual scores produce more misinformation.

We calculate user $u$'s sharing behaviors along the political $p_l(u)$ and factual $f_l(u)$ dimensions using Eqs~\ref{eq:pol_lean} and \ref{eq:fact_lean} respectively.  We denote the set of pay-level domains shared by user $u$ as $D(u)$. These include only the domains appearing in $u$'s original tweets. Functions $\Pi(d)$ and $\Phi(d)$ return the political and factual polarity of each domain $d$. 

\begin{equation}
    p_l(u) = \frac{1}{|D(u)|}\sum_{d \in D(u)} \Pi(d)
    \label{eq:pol_lean}
\end{equation}

\begin{equation}
    f_l(u) = \frac{1}{|D(u)|}\sum_{d \in D(u)} \Phi(d)
    \label{eq:fact_lean}
\end{equation}

% \begin{equation}
%     p_s(u) = \frac{1}{|E(u)|}\sum_{d \in E(u)} \Pi(d)
%     \label{eq:pol_lean}
% \end{equation}

% \begin{equation}
%     f_s(u) = \frac{1}{|E(u)|}\sum_{d \in E(u)} \Phi(d)
%     \label{eq:fact_lean}
% \end{equation}

%After filtering out users who shared fewer than three PLDs with political and factual polarity, we are left with a little over 350K users. 
Fig.~\ref{fig:domain-scores} shows the distribution of user leanings along the political and factual dimensions. The political distribution is skewed to liberal domains, potentially indicating a bias in the Covid-19 data. Similarly, individual factual leanings are skewed towards factuality, and there are relatively few users sharing misinformation or low-factuality content.

%\subsubsection{Quantifying exposure to information}
\subsubsection{Neighborhood Polarization (Exposure)}
Understanding the polarization of information people \textit{see} online is challenging for several reasons. On Twitter, as on other social media platforms, users subscribe to follow accounts of other users to see the content they post. However, the follower graph is usually not available nor is it feasible to reconstruct it from the available APIs.  Even when the follower graph is known, platform's personalization algorithms may select only a subset of the messages posted by friends, i.e., the accounts the user follows, in the user's timeline~\cite{bakshy2015exposure}. This can dramatically change not just the number but also the nature of the information people see~\cite{nathan2020bias,chen2020neutral}. %Beyond this, what posts users see depends on when they log in and how active their friends are.

As a proxy of the follower graph, we use the retweet graph, creating links to the accounts the user retweets. We consider the retweeted accounts as friends whose activity the user sees. We extract tweets and retweets generated by friends, extract PLDs and filter out ones that do not have a political or factual polarity. In contrast to previous works~\cite{Cinellie2023301118,garimella2018political,nikolovright}, however, which measure polarization of a user's neighborhood by averaging over friends' individual political leanings, we aggregate over all messages  posted by friends and calculate political and factual scores of aggregated tweets. This approach factors in the large variation in friend activity: an active friend who posts many messages will have a bigger effect on the user's information exposure than a less active friend.

Information exposure scores along political ($p_e(u)$)  and factual ($f_e(u)$) dimensions are calculated using Equations~\ref{eq:pol_lean} and \ref{eq:fact_lean}, but now the set of pay-level domains $D(u)$ corresponds to all domains user $u$ sees, which we construct by aggregating over all PLDs shared by $u$'s friends.

% The exposure scores are calculated as shown in Equations \ref{eq:pol_exp} and \ref{eq:fact_exp}. $p_e(u)$ and $f_e(u)$ represents the political and factual scores of the information user $u$ sees. We denote the set of pay-level domains the user $u$ is exposed to as $D_e(u)$. As before, $\Pi(d)$ and $\Phi(d)$ return the political and factual polarity of domain $d$. 

% \begin{equation}
%     p_e(u) = \frac{1}{|D_e(u)|}\sum_{d \in D_e(u)} \Pi(d)
%     \label{eq:pol_exp}
% \end{equation}

% \begin{equation}
%     f_e(u) = \frac{1}{|D_e(u)|}\sum_{d \in D_e(u)} \Phi(d)
%     \label{eq:fact_exp}
% \end{equation}

\begin{figure}[tb]
    \centering
    \includegraphics[width=0.9\linewidth]{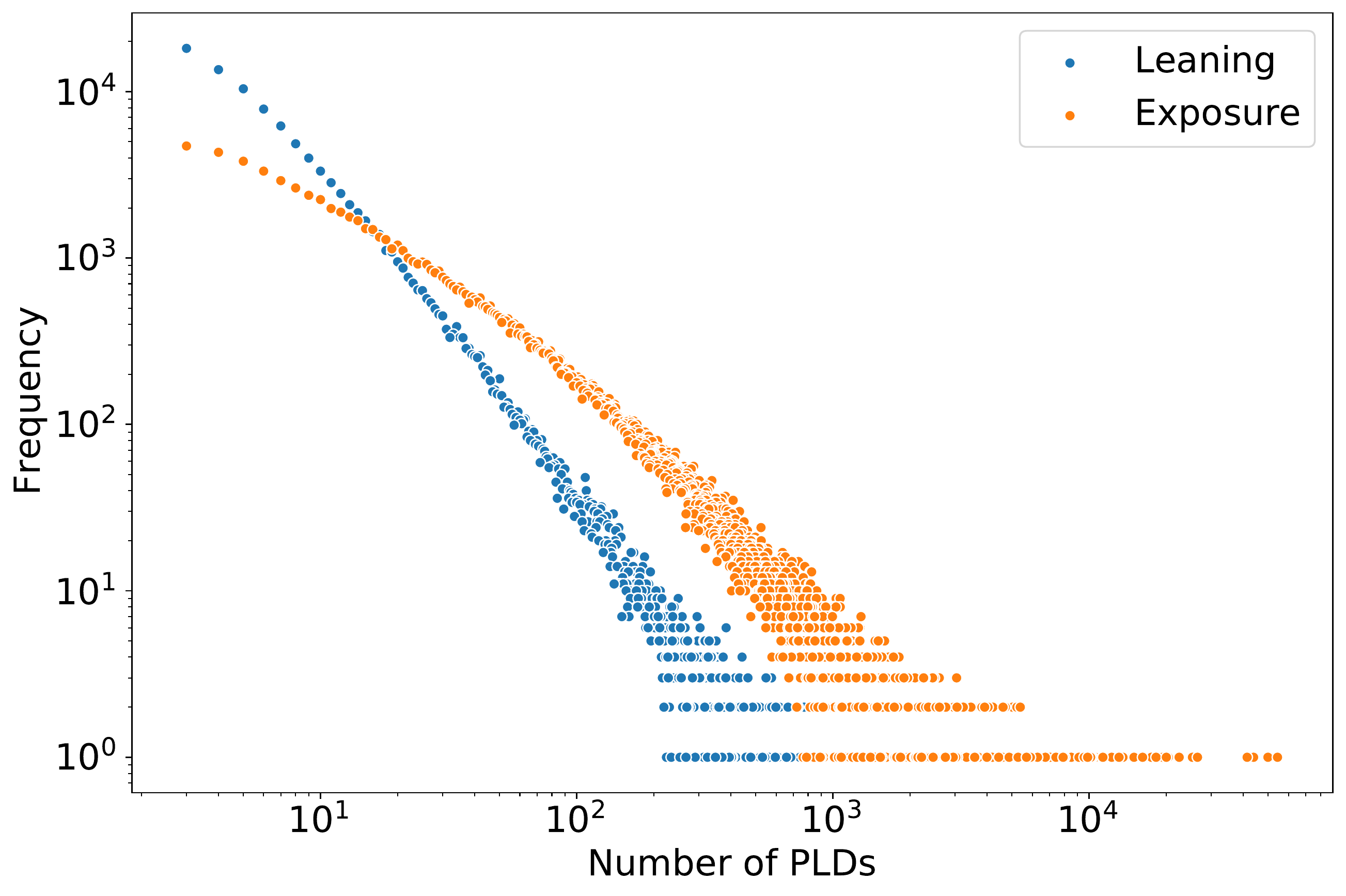}
    \caption{Distribution of number of PLDs users generate (shown in blue) and are exposed to (shown in orange).}
    \label{fig:pld_dist}
\end{figure}

After filtering out users who share or see two or fewer PLDs with political and factual polarity, we are left with a little over 350K users. 
Fig.~\ref{fig:pld_dist} shows the distribution of the number of pay-level domains users share in their posts, as well as the distribution of the number of PLDs users see. The difference between the two distributions suggests that some  domains are seen much more than they are shared, likely because they are shared by influential accounts with many followers. In this paper we study the relationship between the polarization of information people see on social media and polarization of information they themselves share.

%amplifying polarized content and misinformation while the other side is filtering/dampening out polarized content and misinformation.
%Rename the axes as amplifying and dampening.

%attention at political extremes - color heatmap figure 8

\section{Results}

\begin{figure}[h!]
    \centering
    \subfigure[Political vs Political Leaning]
    {\includegraphics[width=\linewidth]{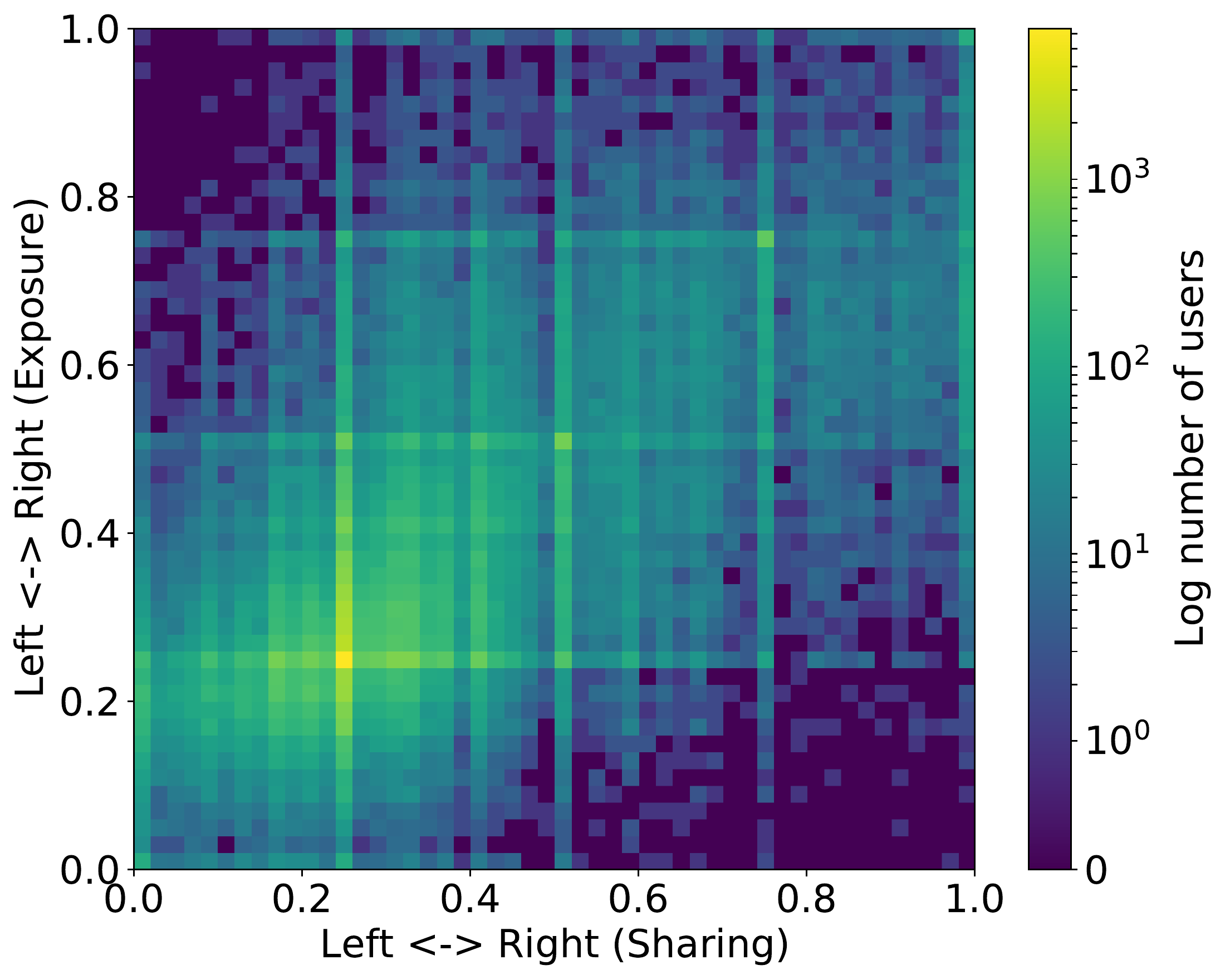}}
    \subfigure[Factual Exposure vs Factual Leaning]
    {\includegraphics[width=\linewidth]{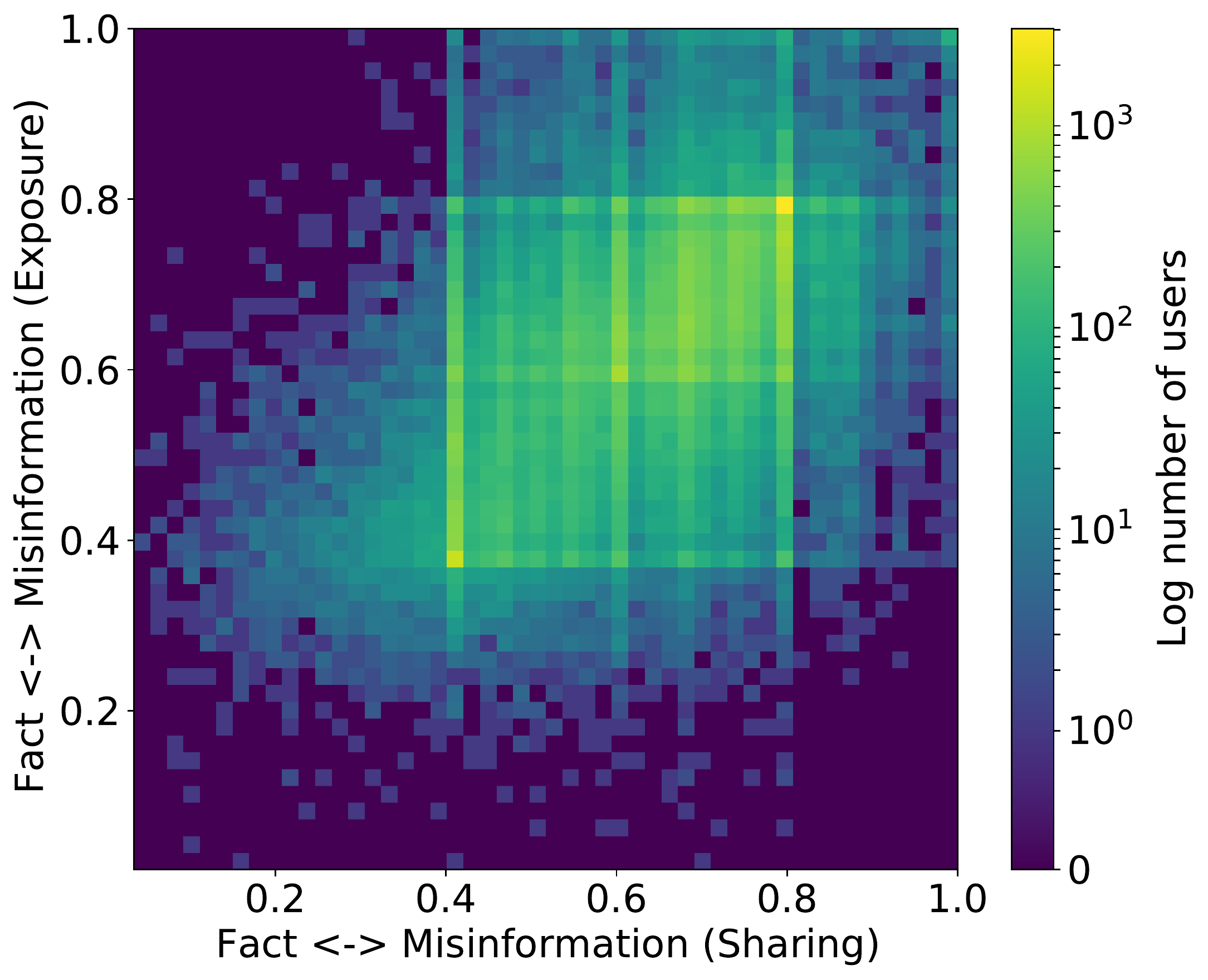}}
    \caption{Ideological echo chambers. Heatmap of (a) political leaning and political exposure ($r=0.61^{***}$) %; political exposure predicts political leaning with $R^2 = 0.611$. 
    and (b) factual leaning and factual exposure ($r=0.50^{***}$). Colors indicate the number of users.}
    %Factual exposure predicts factual leaning with  $R^2 = 0.542$.textbf{SWITCH X,Y}}
    \label{fig:echo_individual}
\end{figure}

We now explore the relationship between individual polarization of users discussing Covid-19 online and the polarization of information  to which they are exposed.

\subsection{Existence of Political and Factual Echo-Chambers}

First, we explore the relationship between individual polarization and polarization of information exposure separately along each dimension.
Fig.~\ref{fig:echo_individual} shows the joint distribution of individual political (Fig.~\ref{fig:echo_individual}(a)) and factual (Fig.~\ref{fig:echo_individual}(b)) leanings and the political (resp. factual) scores of the information they are exposed to by friends in their retweet neighborhood. %by visualizing their political and factual leaning and exposure domain scores for 345K users. 
The high density along the diagonal confirms the existence of echo chambers: many users are linked to friends who expose them to ideologically congruent information. The correlation between individual leanings and information exposure scores along the political and factual dimensions are $0.61\ (p<0.001)$ and $0.5\ (p<0.001)$ respectively. There are no partisan asymmetries in the political echo chambers (Fig.~\ref{fig:echo_individual}(a)), as both liberal and conservative users are exposed to a similar variety of political content. There is some asymmetry in factual information echo chambers (Fig~\ref{fig:echo_individual}(b)), since there is much lower density of users in the misinformation bubble.
Unlike previous works, e.g., \cite{Cinellie2023301118}, the echo chambers we observe are more diffuse, with users linked to friends with more variable ideologies. This is because previous works calculate the average polarization of friends, which gives equal weight to friends who share a lot or little information, while we aggregate in- formation shared by all friends when measuring polarization.

%We then leverage the two quantified dimensions of interest - politics and content factuality to study the relationship between them and consequently highlight the multidimensionality of echo chambers. Finally, we discuss how political ideology related to a user's propensity to misinformation.

\subsection{Polarization is Multi-dimensional}

\begin{figure}[bht]
    \centering
    % \subfigure[Political Leaning vs Factual Leaning]{\includegraphics[width=1.0\linewidth]{./figures/user_production_scores.pdf}}
    \includegraphics[width=\linewidth]{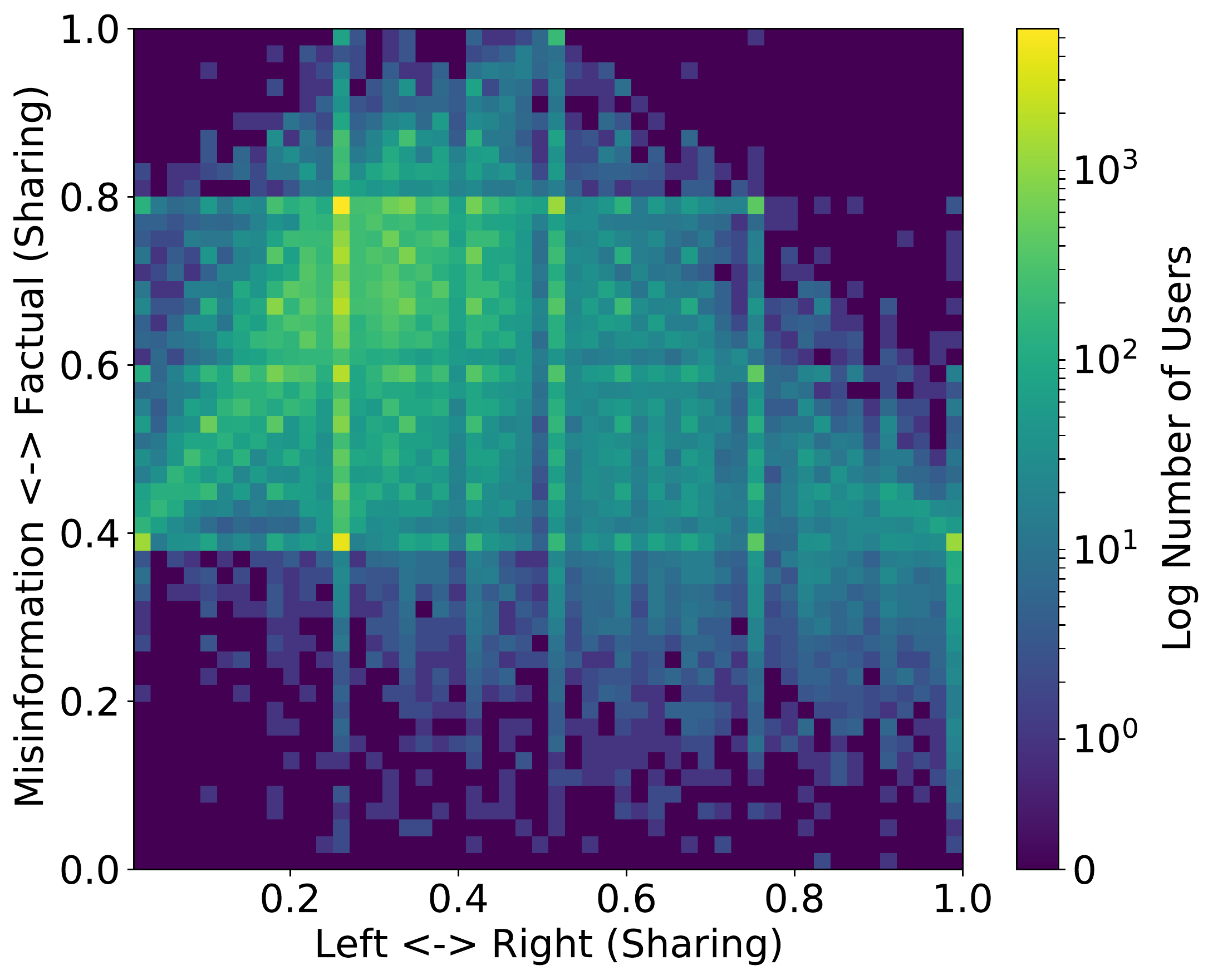}
    \caption{Relationship between individual political and factual leanings. Color represents number of users. Pearson's correlation $r=-0.198\ (p<0.001)$.}
    \label{fig:prod_scores_individual}
\end{figure}

Previous research has shown that political and factual dimensions of information people share online are correlated: conservatives share misinformation to a greater degree than liberals~\cite{vousoghi,Grinberg374,nikolovright}, and they also tend to share more anti-science sources~\cite{rao2021political}. Our results are consistent with these findings.  Fig.~\ref{fig:prod_scores_individual} shows the distribution of user scores in the political-factual space. There is a strong negative correlation ($-0.198,\ p<0.001$) between user leanings along the two dimensions: users sharing more conservative domains are more likely to share misinformation. However, the large variance masks more nuanced positions. For example, the bright line in the upper-left quadrant shows a phenomenon also observed by \cite{nikolovright} that more extreme liberals have a greater propensity to share misinformation.

%Heatmap of exposure]]

\begin{figure}[tbh]
    \centering
    \subfigure[Factual topics]
    {\includegraphics[width=0.9\linewidth]{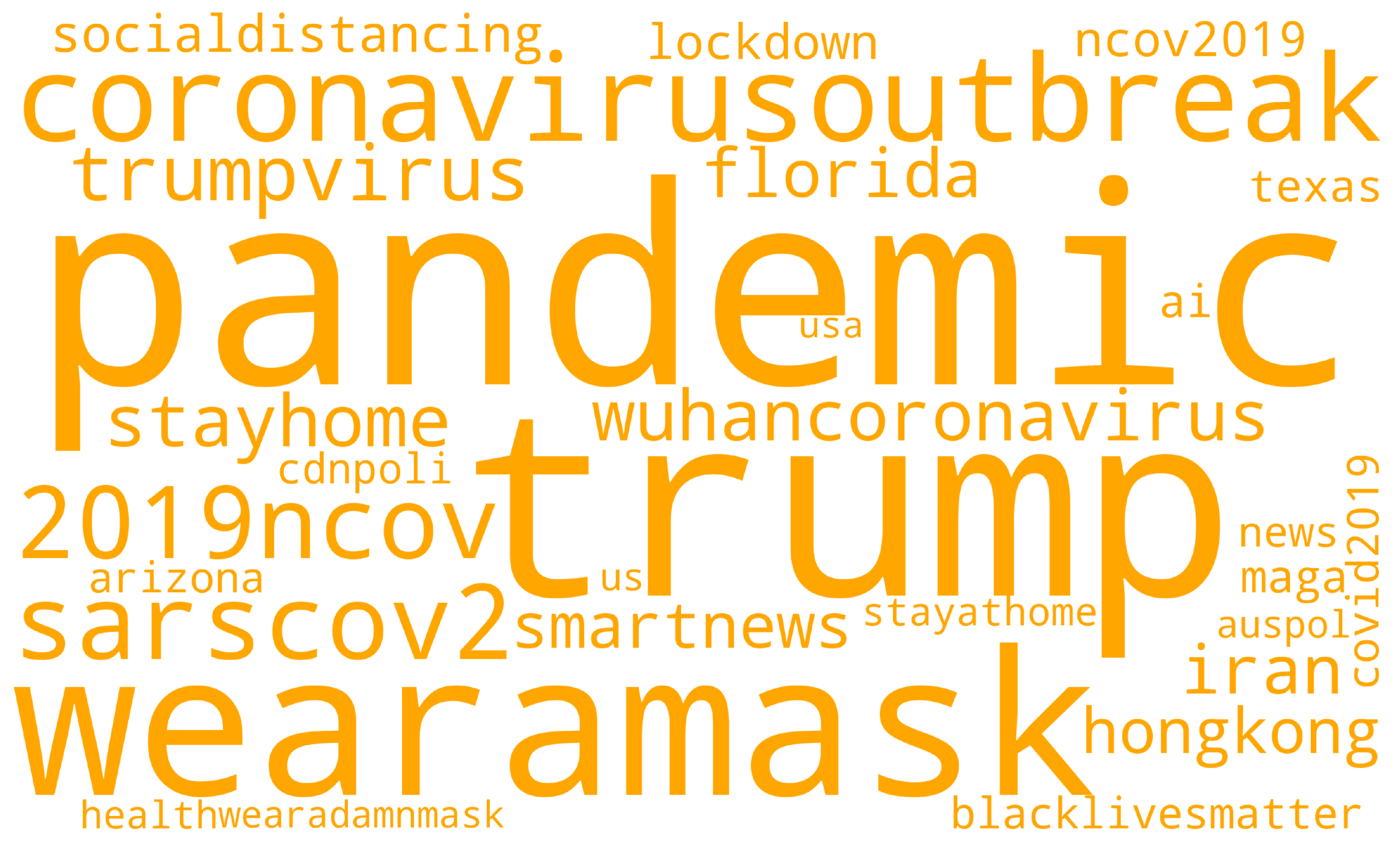}}
    \subfigure[Misinformation topics]
    {\includegraphics[width=0.9\linewidth]{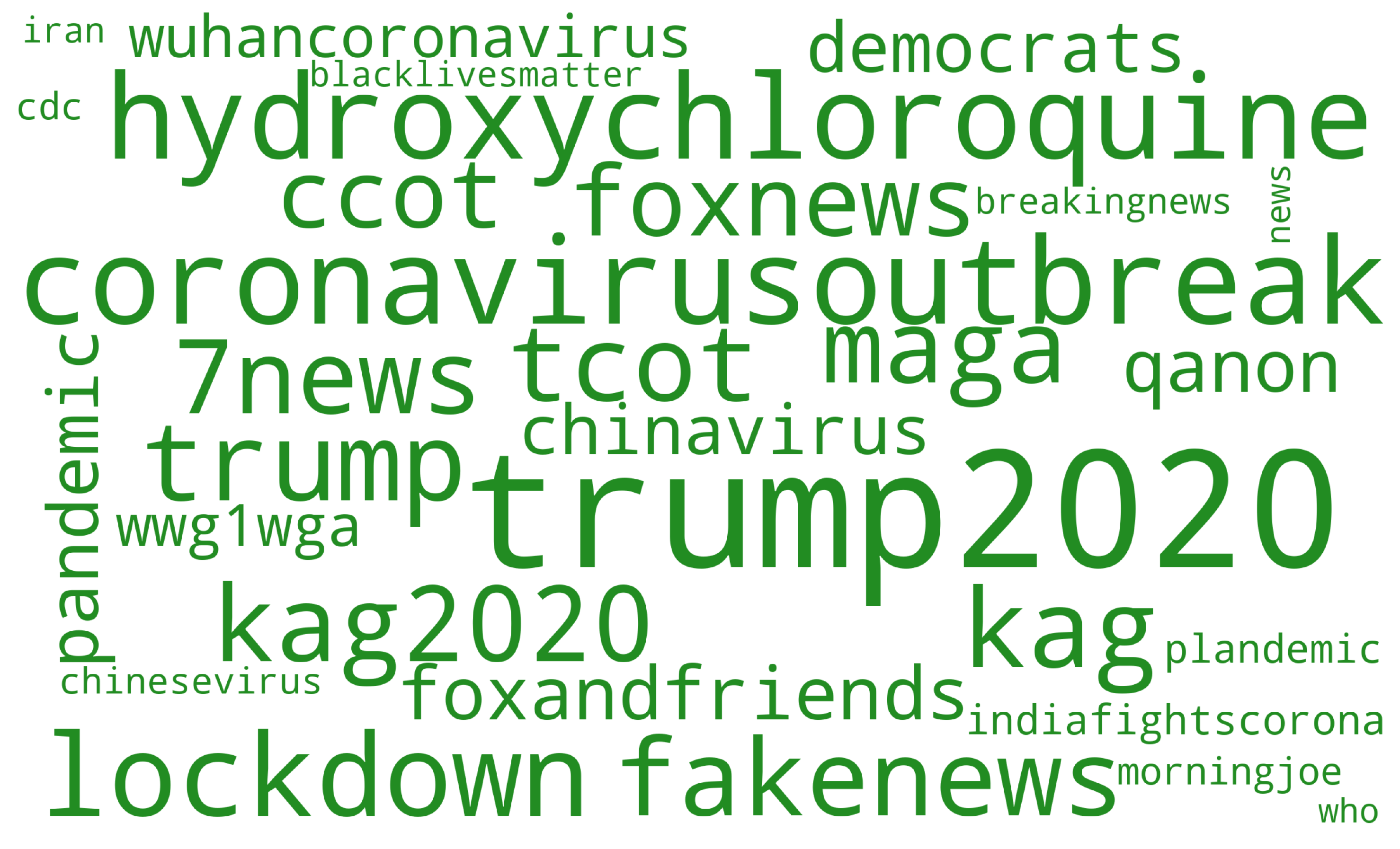}}
    % \subfigure[Liberal topics]{\includegraphics[width=1\linewidth]{./figures/left_wordcloud.pdf}}
    % \subfigure[Conservative topics]{\includegraphics[width=1\linewidth]{./figures/right_wordcloud.pdf}}
    \caption{Word cloud of popular hashtags frequently used by (a) factual users ($f_p \geq 0.6$) and (b) users with high propensity for misinformation  ($f_p \leq 0.4$). % (a) liberal ($p_l \leq 0.4$) and (b) conservative users ($p_l \geq 0.6$)
    }.
    \label{fig:wordclouds}
\end{figure}

% Figure~\ref{fig:wordclouds} shows the wordclouds of popular hashtags used by liberal and conservative users.  While liberals post messages on health topics, such as ``pandemic'', ``wearamask'', ``stayhome'', conservative users are preoccupied with politics (``trump2020'', ``kag2020'', ``democrats'', ``maga''), and conspiracies (``qanon'', ``wwg1wga''). Interestingly, conservatives discuss media to a greater extent than liberals, mentioning topics like ``foxnews'', ``7news'', ``9news'', ``foxandfriends'', ``morningjoe'', ``oann'' and ``fakenews''.
Fig.~\ref{fig:wordclouds} contrasts popular topics (hashtags) discussed by people sharing factual information and misinformation.  While factual people post messages on health topics, such as ``pandemic'', ``wearamask'', ``stayhome'', people sharing misinformation are preoccupied with politics (``trump2020'', ``kag2020'', ``democrats'', ``maga''), conspiracies (``plandemic'', ``qanon'', ``wwg1wga''). Interestingly, these users also mention media to a much greater extent, using topics like ``foxnews'', ``7news'', ``foxandfriends'', ``morningjoe'', and ``fakenews''. This may suggest the greater role that media plays in agenda-setting for people vulnerable to misinformation. Also, unlike factual users, people spreading misinformation also discuss unproven cures, like ``hydroxychloroquine''.

\subsection{Partisan Asymmetries in Exposure to Misinformation}

\begin{figure*}[tbh]
    \centering
    \begin{tabular}{|cc|}
    \hline
    \multicolumn{2}{|c|}{Information exposure within multi-dimensional echo chambers}\\ 
%    \subfigure[Political Leaning vs Political Exposure]
    {\includegraphics[width=0.4\linewidth]{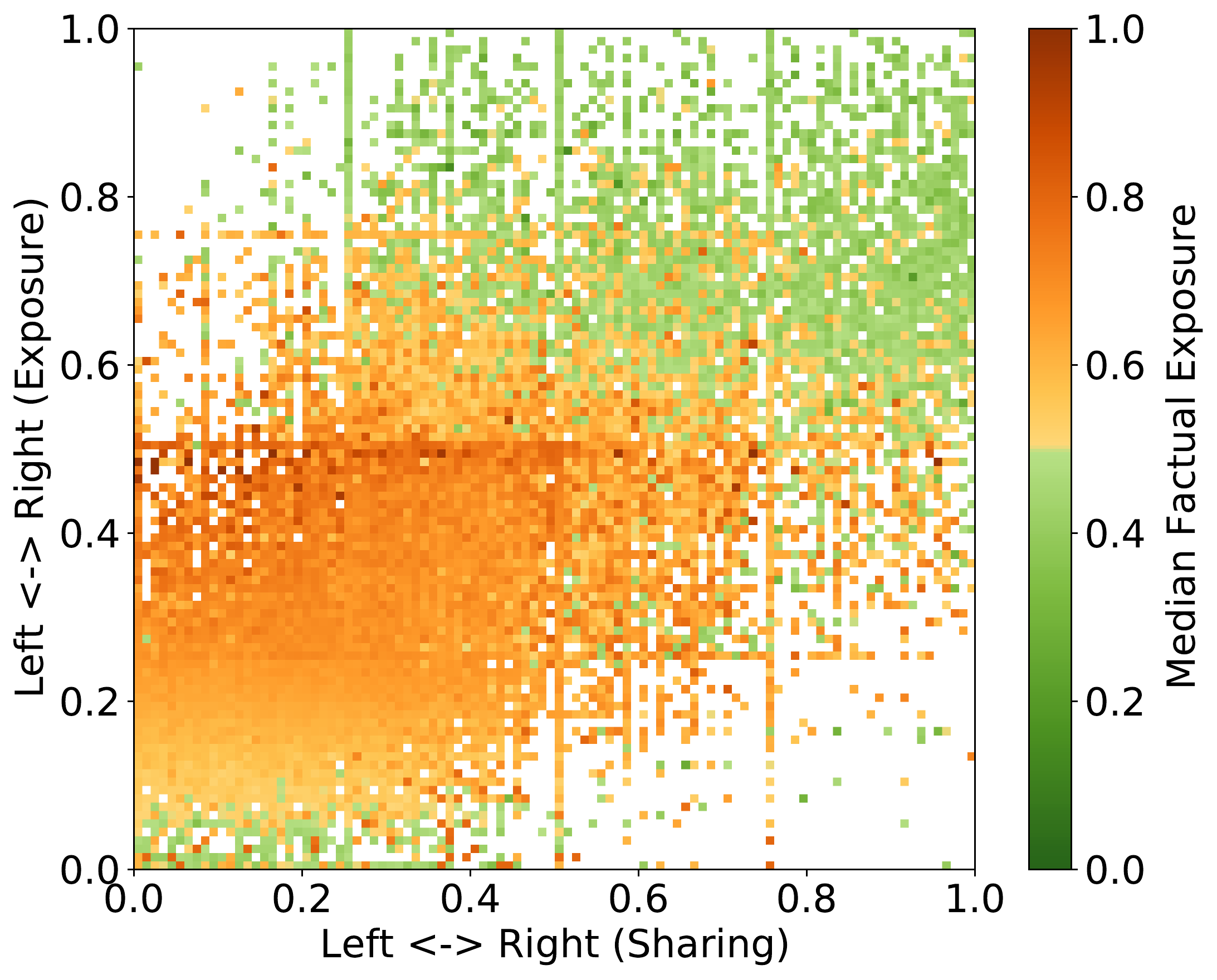}} &
    %\subfigure[Factual Leaning vs Factual Exposure]
    {\includegraphics[width=0.4\linewidth]{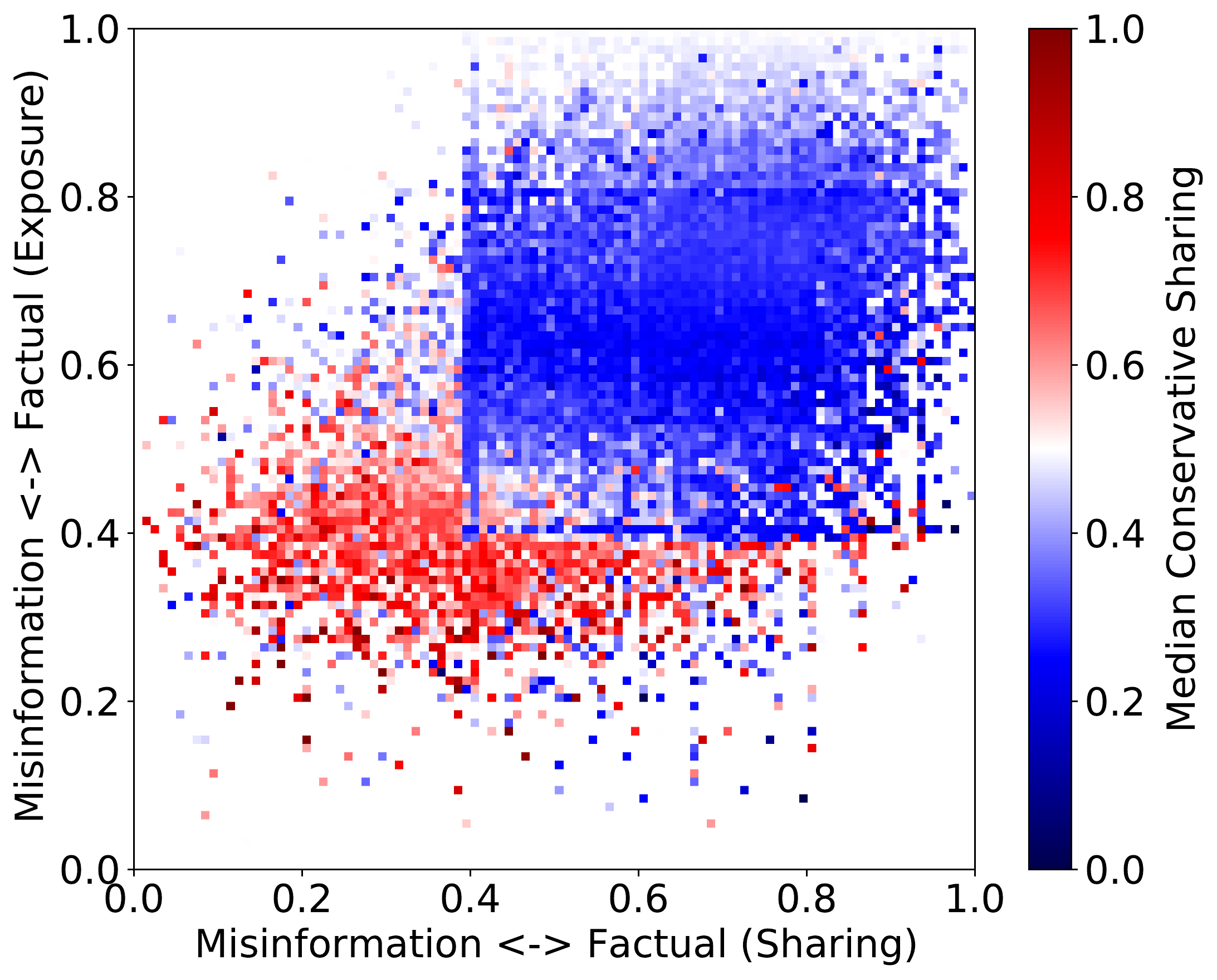}} \\ %\vspace{-2em}
    (a) Political Exposure $p_e$ vs Political Sharing $p_l$ &
    (b) Factual Exposure $f_e$ vs Factual Sharing $f_l$ \\ \hline
    \multicolumn{2}{|c|}{Individual polarization within multi-dimensional echo chambers} \\ 
    %\subfigure[Political Leaning vs Political Exposure]
    {\includegraphics[width=0.4\linewidth]{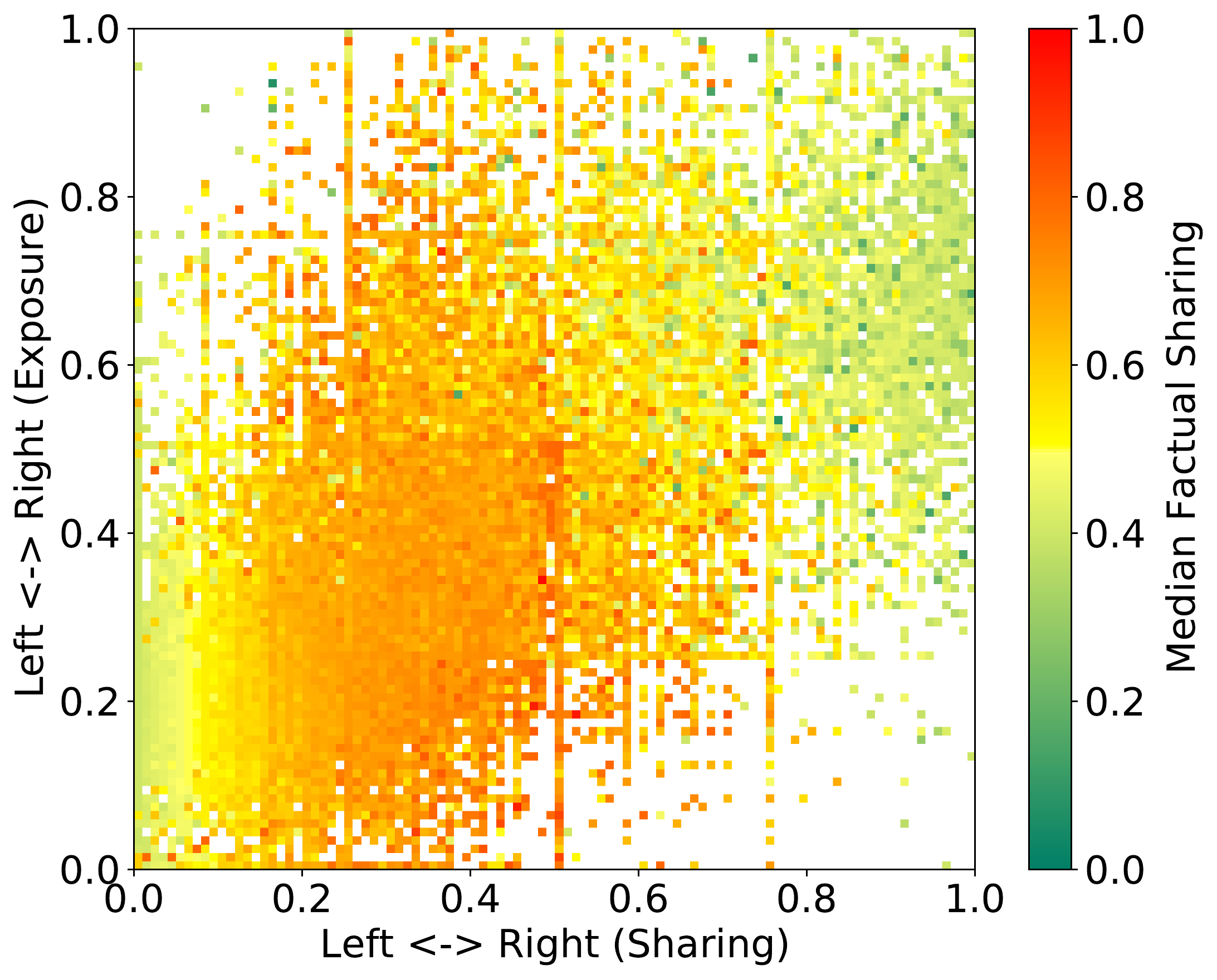}} &
    %\subfigure[Factual Leaning vs Factual Exposure]
    {\includegraphics[width=0.4\linewidth]{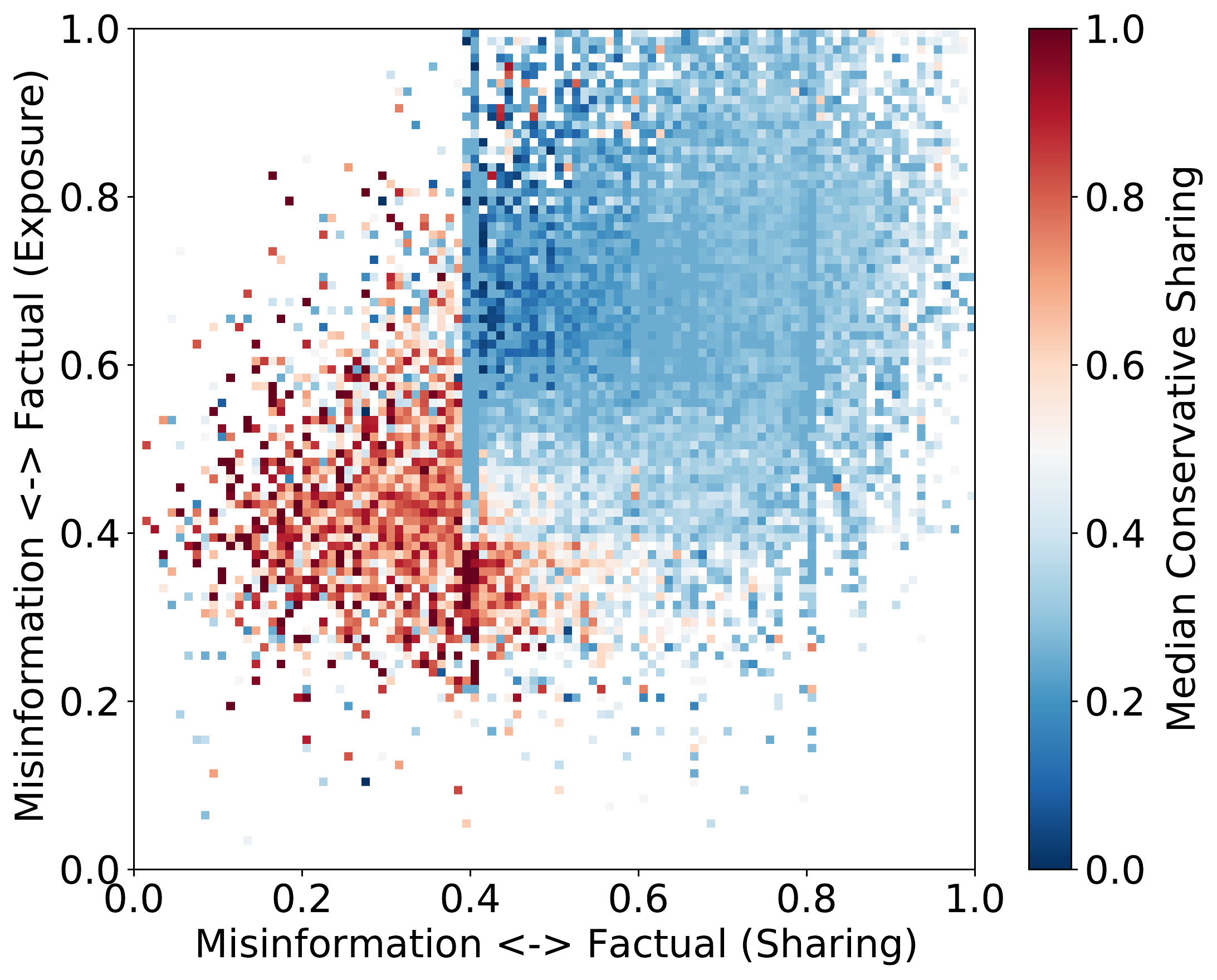}} 
    \\ %\vspace{-4em}
    (c) Political Exposure $p_e$ vs Political Sharing $p_l$ &
    (d) Factual Exposure $f_e$ vs Factual Sharing $f_l$
    \\ \hline
    \end{tabular}

\caption{Multi-dimensional polarization and information exposure within echo chambers. (a) Color indicates the median factual leaning score in each bin. In general, as users produce more conservative content while being exposed to more conservative content, they have a higher propensity to misinformation. Low content factuality is also seen along the liberal extreme where users produce far-left content. (b) Color indicates the median political leaning score in each bin. Generally, as users generate more misinformation while being exposed to low factual content, they have a higher propensity to political conservatism.}%(c) Color indicates the median factual exposure score in each bin. (d) Color indicates the median political exposure score in each bin.}
    \label{fig:echo_interplay}
\end{figure*}

How does the interaction between partisanship and factuality affect what information the  users are exposed to within their echo chambers and, in turn, what information they share? Do people effectively filter out misinformation they see by selectively sharing more factual content? Or do they amplify misinformation by selectively sharing fewer factual domains than what they are exposed to?  

Fig.~\ref{fig:echo_interplay} visualizes user exposure to multi-dimensional information within the echo chambers. The top row shows user exposure to political and factual information as a function of user political (Fig.~\ref{fig:echo_interplay}(a)) and factual (Fig.~\ref{fig:echo_interplay}(b)) leanings. Note the neighborhood exposure vs leaning space is the same shown in Fig.~\ref{fig:echo_individual}(a) and (b) for respectively. The color in each plot shows the median exposure score.  There are several regions of interest in Fig.~\ref{fig:echo_interplay}(a). Liberal users ($p_l < 0.5$) who are exposed to politically moderate content ($p_e \approx 0.5$) see the most factual information (dark orange). Liberals ($p_l < 0.5$) who are exposed to liberal content ($p_e < 0.5$) generally see more factual (orange) information, although as their exposure becomes more partisan, the share of misinformation they see grows. Those exposed to extreme left content ($p_e \approx 0$) see more misinformation (green hue). As liberals become more exposed to conservative content ($p_e \rightarrow 1$) they see more and more misinformation %(perhaps their goal is to troll conservatives). 
The same is not true of conservatives: conservative users ($p_l>0.5$) who are exposed to right-wing information ($p_e>0.5$) tend to see more misinformation; however, as long as they are not too conservative, exposure to liberal information ($p_e<0.5$) allows them to receive more factual information. Unlike liberals, exposure to politically moderate content ($p_e \approx 0.5$) does not promote factual information among conservatives.

Trends within misinformation echo chambers (Fig.~\ref{fig:echo_interplay}(b)) tell a similar story. Users who generate more misinformation ($f_l<0.4$) and are exposed to misinformation ($f_e<0.4$) tend to see more conservative content (red), although those who are exposed to more factual content ($f_e \rightarrow 1$) see more liberal information (blue dots). Among people sharing factual information ($f_l>0.6$), those who are exposed to more factual information ($f_e \rightarrow 1$) tend to see politically moderate content (white).
The box outline is an artifact of domain polarity scores. MBFC classifies many information sources as ``mixed'' (0.4), leading to an overabundance of points near that value.

The bottom row of Fig.~\ref{fig:echo_interplay} visualizes multi-dimensional polarization within the echo chambers. Again, the neighborhood exposure vs leaning space is the same as the row above, but the color in each plot shows user polarization or leaning along the alternate dimensions. Fig.~\ref{fig:echo_interplay}(c) shows that as partisanship becomes more extreme ($p_l \rightarrow 0$ or $p_l \rightarrow 1$), people are more likely to share misinformation (green). Interestingly, this trend does not strongly depend on partisanship of their exposure ($p_e$). Overall, liberals ($p_l < 0.5$) share more factual information, although those who are more moderate ($p_l \approx 0.5$) tend to share more misinformation (yellow/green) as they are exposed to more conservative content ($p_e \rightarrow 1$). As shown in Fig.~\ref{fig:echo_interplay}(d), misinformation-prone users ($f_l<0.4$) tend to post more hardline conservative content (darker red) as they share more misinformation ($f_l \rightarrow 0$) regardless of their exposure; however, those who are most exposed to misinformation ($f_e<0.2$) tend to share more liberal views (blue dots). This is not true for factual users, who tend to share liberal content (blue) regardless of the factuality of their exposure ($f_e$).

%\subsection{Polarization in Multi-dimensional Echo Chambers}
\subsection{Hardline Partisans Amplify Misinformation} % Excess Misinfo

The off-diagonal elements in the echo chamber plots in Fig.~\ref{fig:echo_individual} suggest that a sizable fraction of social media users share information that is more polarized and less factual than what they are exposed to, and an equally large number share information that is more factual than what they are exposed to. In other words, some people filter out misinformation from the information ecosystem, while others amplify it. To better understand how the patterns in the interaction between political and factual dimensions affect how people react to exposure, we define two quantities. 

\begin{eqnarray}
    \Delta_f(u) &=& f_l(u) - f_e(u) \\
        \label{eq:net-factuality}
    \Delta_p(u) &=& |p_l(u)-0.5| - |p_e(u)-0.5|
    \label{eq:net-partisanship}
\end{eqnarray}

Equation~\ref{eq:net-factuality} simply gives excess factuality (amplified factuality) for a given user, i.e., how much more factual content the user shares relative to their exposure.
Equation~\ref{eq:net-partisanship} measures excess partisanship (amplified partisanship), i.e., the relative partisanship of the content the user shares compared to their exposure. Note that we had transformed polarization scores so that instead of partisanship, they measure the degree of political moderacy or  extremism regardless of its ideological label.

%  We propose two new measures:
% %In order to quantify political moderacy (and, extremism), both in terms of leaning and exposure, we propose two new measures - 
% moderacy of political leaning $(m_l)$ and moderacy of political exposures $(m_e)$. We obtain these scores by applying a simple transformation to the leaning ($p_l$) and exposure scores ($p_e$) as shown in Equations \ref{eq:pol_extremism_l} and \ref{eq:pol_extremism_e} respectively. $m_l$ and $m_e$ range between $[0.5,1.0]$ with, $0.5$ being the most moderate and $1.0$ being the most extreme.

% \begin{equation}
%     m_l=\begin{dcases*}
% 			1-p_l  & $if\,\, p_l < 0.5,$\\
%             p_l & $otherwise$
% 		 \end{dcases*}
% 	\label{eq:pol_extremism_l}
% \end{equation}

% \begin{equation}
%     m_e=\begin{dcases*}
% 			1-p_e  & $if\,\, p_e < 0.5,$\\
%             p_e & $otherwise$
% 		 \end{dcases*}
% 	\label{eq:pol_extremism_e}
% \end{equation}

% As a means to understand the effects of exposure to moderate (or, hardline) content on the moderacy of political leaning, we measure net-moderacy $N_m$ for user $u$ as,

% \begin{equation}
%     N_m(u) = m_l(u) - m_e(u).
%     \label{eq:net-moderacy}
% \end{equation}

\begin{figure}[tbh]
    \centering
    \includegraphics[width=1\linewidth]{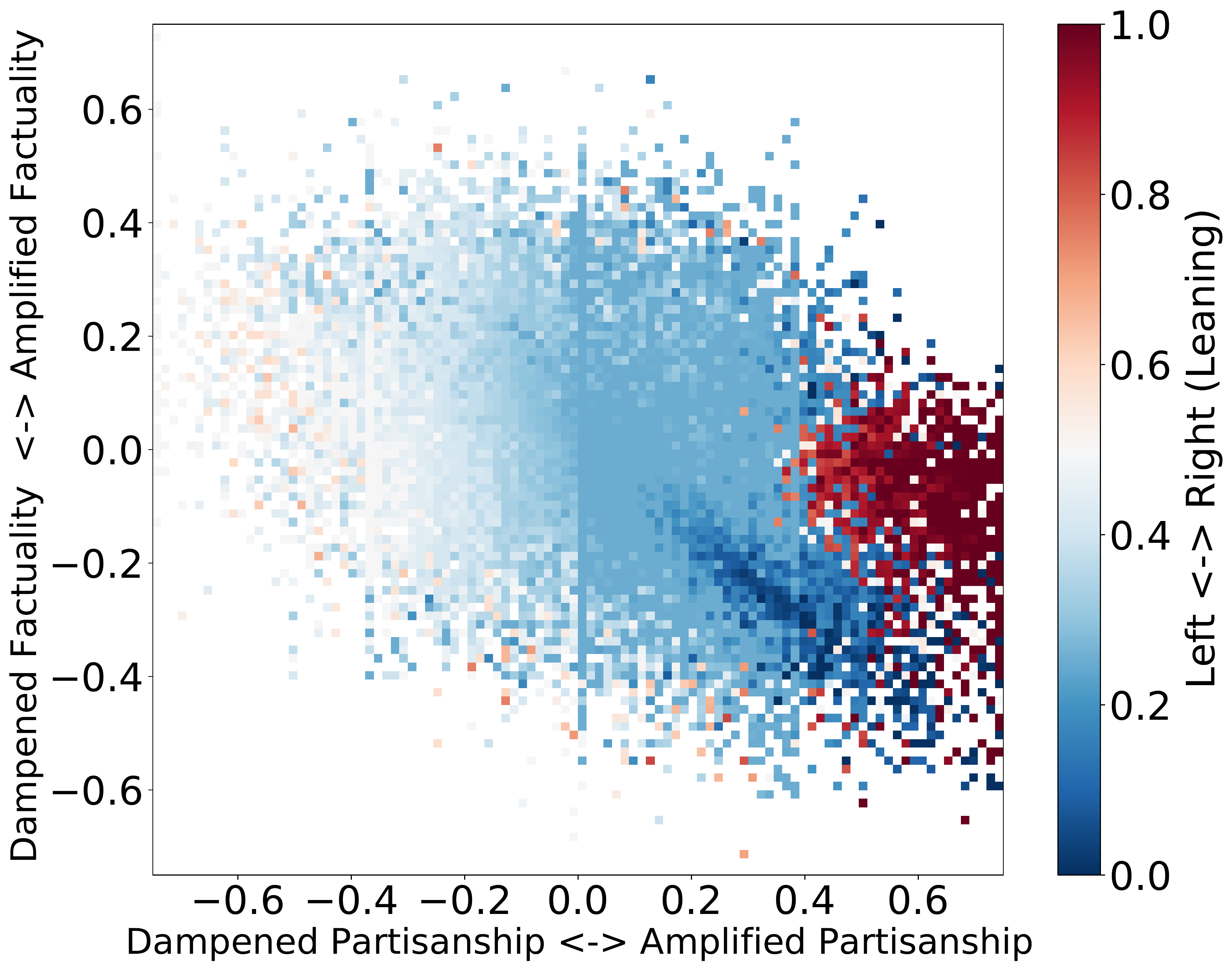}
    \caption{Excess factuality $\Delta_f$ vs Excess partisanship $\Delta_p$}
    \label{fig:hardline-vs-factuality}
\end{figure}

\begin{figure}[tbh]
    \centering
    \includegraphics[width=1\linewidth]{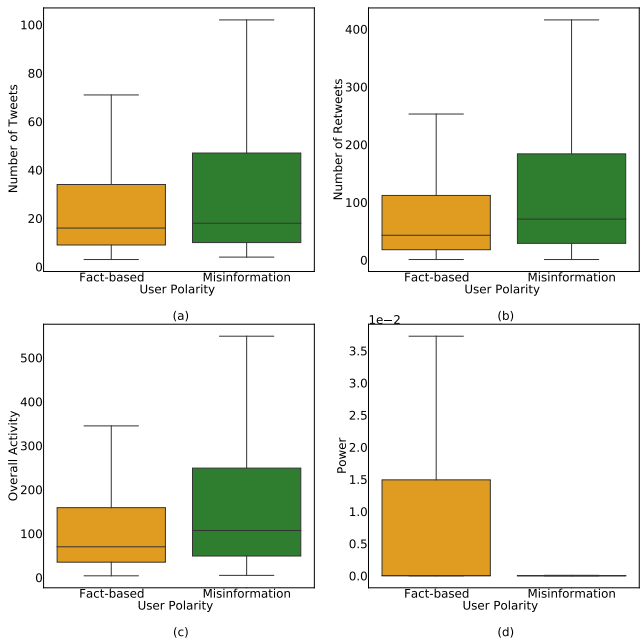}
     \caption{Boxplots comparing activity and power of factual ($f_l \geq 0.6$) and misinformative users ($f_l \leq 0.4$). While we notice that misinformative users are more active both in terms of number of tweets and retweets generated, they are retweeted less com- pared to factual users. Subsequently, the ratio of retweets received to overall activity is significantly lower for misinformative users than factual ones.}
    \label{fig:activity-boxplots}
\end{figure}

\begin{table*}[tbh]
\centering
    \footnotesize
\renewcommand{\arraystretch}{1.5}
\begin{tabular}{p{2.5cm} p{2.5cm} p{3.5 cm} p{5.7 cm} p{2cm}}\hline
  \textbf{Metric $(\theta)$} & \textbf{Factual ($\mu(\theta)_F$)} & \textbf{Misinformation ($\mu(\theta)_M$)} & \textbf{Hypotheses} & \textbf{t-statistic} \\
    \hline
\multirow{2}{*}{$T$} & \multirow{2}{*}{$38.93$} &\multirow{2}{*}{$\textbf{57.01}$} & $H_0:\mu(log(T))_{F}=\mu(T)_{log(M)}$ & \multirow{2}{*}{$\textbf{16.75}^{***}$}\\
 & & &$H_a:\mu(log(T))_{F} \neq \mu(log(T))_{M}$& \\
 
  \multirow{2}{*}{$RT$}& \multirow{2}{*}{$120.69$} &\multirow{2}{*}{$\textbf{168.28}$} & $H_0:\mu(log(RT))_{M}\leq\mu(log(RT))_{F}$ & \multirow{2}{*}{$\textbf{27.42}^{***}$}\\
  & & &\textbf{$H_a:\mu(log(RT))_{M} > \mu(log(RT))_{F}$}& \\
 
  \multirow{2}{*}{$A = T+RT$} & \multirow{2}{*}{$159.63$} &\multirow{2}{*}{$\textbf{225.29}$} & $H_0:\mu(log(A))_{M}\leq\mu(log(A))_{F}$ & \multirow{2}{*}{$\textbf{28.89}^{***}$} \\
    & & &\textbf{$H_a:\mu(log(A))_{M} > \mu(log(A))_{F}$}& \\
    
  \multirow{2}{*}{$R$} & \multirow{2}{*}{$\textbf{61.97}$} &\multirow{2}{*}{$48.82$} & $H_0:\mu(log(R))_{F}\leq\mu(log(R))_{M}$ & \multirow{2}{*}{$\textbf{25.64}^{***}$}\\
    & & &\textbf{$H_a:\mu(log(R))_{F} > \mu(log(R))_{M}$}& \\
    
  \multirow{2}{*}{$P = R/A$} & \multirow{2}{*}{$\textbf{0.57}$} &\multirow{2}{*}{$0.17$} & $H_0:\mu(log(P))_{F}\leq\mu(log(P))_{M}$ & \multirow{2}{*}{$\textbf{32.86}^{***}$}\\
    & & &\textbf{$H_a:\mu(log(P))_{F} > \mu(log(P))_{M}$}& \\
\hline

\end{tabular}
\caption{Results of hypothesis testing for difference in means between the two groups of users along the factuality dimension for various metrics. Factual users $(F)$ have high factuality scores ($f_l \geq 0.6$) while misinformation users $(M)$ have low scores ($f_l \leq 0.4$). Metrics include: number of tweets $(T)$ and retweets $(RT)$ generated by the user, the overall activity $(A)$, number times the user is  retweeted $(R)$ and retweet power $(P)$ which is the ratio of number of times retweeted and activity. We performed $t$-tests to  assess the statistical significance of difference between the two distributions after log transforming the variables. %In order to satisfy Gaussian assumptions of t-tests, we project the metrics onto the logarithmic (base-10) scale. 
$^{***}$ denotes a statistically significant difference between the means of the two distributions with p-value $<0.001$.}
\label{tab:fact_activity}
\end{table*}

%We are now in a position to answer how users filtering out moderate content relates to users filtering out factual content and leverage the net-factuality scores calculated earlier to study this relationship. Figure \ref{fig:hardline-vs-factuality} visualizes this relationship. With a Pearson's correlation of $ r = -0.33$, statistically significant at $p < 0.001$, we see that users who filter out moderate content seemingly generate less factual content than their exposures. The color densities also point to the politically extremist leaning of this region. On the other hand, we see that users who filter out hardline content, mostly generate higher factual content and are politically moderate.

Fig.\ref{fig:hardline-vs-factuality} shows the joint distribution of excess partisanship $\Delta_p$ and excess factuality $\Delta_f$. The negative correlation (Pearson's correlation $ r = -0.38, $ $p < 0.001$) between the two dimensions suggests that not only do politically hardline social media users (regardless of whether they are liberal or conservative) have a higher propensity for misinformation, but users who amplify politically polarized content also amplify misinformation. The color shows partisanship. Interestingly, both hardline conservatives and hardline liberals are active in amplifying partisanship $\Delta_p>0$ and misinformation $\Delta_f<0$, with liberals playing a more active role in amplifying misinformation (dampening factuality). On the other hand, users who are less partisan than their neighborhood ($\Delta_p < 0$) also share more factual information than what they are exposed to ($\Delta_f>0$). By filtering out misinformation, such users play an important role in the information ecosystem. They also tend to be politically moderate.

\subsection{Less Attention to Misinformation} %Factual Content}

The presence of users generating low factuality content raises  questions about their activity and the subsequent attention they garner. Are users sharing misinformation more active than users sharing more factual content? Does aggressive content generation correlate with more attention? These are some of the questions that become imperative to assess the ravages of misinformation on social media.

We define a user's overall activity as the sum of the tweets $T$ and retweets $RT$ they generate $A(u)=T(u)+RT(u)$. In order to quantify the attention the user $u$ receives in response to their activity, we define \textit{retweet power}  $P(u)$ as the ratio of number of times $u$ is retweeted $R$ and their overall activity:  

% \begin{equation}
%     A(u) = T(u) + RT(u)
%     \label{eq:overall_activity}
% \end{equation}

\begin{equation}
    P(u) = \frac{R(u)}{A(u)} = \frac{R(u)}{[T(u)+RT(u)]}
    \label{eq:power}
\end{equation}

Boxplots in Fig.~\ref{fig:activity-boxplots} visualize the differences in tweet and retweet activity of factual $(f_l \geq 0.6)$ and misinformation $(f_l \leq 0.4)$ users. To assess the significance of differences between the two groups, we use the \textit{Student's $t$-test}. This parametric test of difference between the means of two groups requires the corresponding distributions to be normal. While our metrics (the number of tweets and retweets) have a skewed distribution, taking a log transform increases normality. Table \ref{tab:fact_activity} details the null and alternate hypotheses used in our $t$-tests.

From Fig.\ref{fig:activity-boxplots} and Table \ref{tab:fact_activity}, we see that misinformative users tweet and retweet more often and have higher overall activity compared to factual users. Statistically significant $t$-statistics for $T$,$RT$ and $A$ in Table \ref{tab:fact_activity} reinforce these findings. 

Despite their increased overall activity, users sharing misinformation are retweeted less often than factual users $(\mu(R_M) < \mu(R_F))$, significant at $p<0.001$ and have considerably lower retweet power $(\mu(P_M) < \mu(P_F))$ at $p < 0.001$ (Fig.\ref{fig:activity-boxplots}(d))  . These findings hint at an increased attention to factual users despite their lower overall activity.

\section{Conclusion}
In this work we study multi-dimensional polarization and echo chambers. We focus on two dimensions of polarization---political and factual---which are assessed by measuring the bias in the pay-level domains that the users tweet. We find that there is strong polarization along both dimensions. To deepen the understanding of the mechanics behind echo chambers, we separate a user's interactions into their exposure (what their friends post) and leaning (what they post). We find a strong correlation between what a user sees and what the post, confirming the presence of echo chambers.

Next, we study the partisan asymmetries of these echo chambers. We find that conservatives are more likely to share misinformation than liberals. Nevertheless, extremely liberal partisanship increases a user's propensity to share misinformation. Moreover, we find that moderate liberals have the highest exposure to factual information. We find that this does not hold for conservatives whose exposure to liberal information yields the most truthful information. Furthermore, for conservatives exposure to politically moderate content does not make them more factual.

Lastly, we look at the relationship between partisan extremism and misinformation. %We hypothesize that users act as either filters or amplifiers of misinformation. 
We find that highly polarized users, who represent hardline partisans on both sides of the political spectrum, are most likely to amplify partisan content and misinformation. However, such users get less attention than the bulk of users in our study who are political moderates who selectively share more factual content. Therefore, such users filter out misinformation.

There are several limitations to this study worth considering. First, we do not know the actual exposures and thus rely on the retweet network as a proxy. Second, there could be factual/pro-science bias in the data due to the way it was collected. More generally, the keyword-based Twitter crawl used to produce this data could omit nuanced subtopics related to Covid-19 discussions. Lastly, our study focuses on users in the United States. This decision was made because of the United States' information environment, and due to the dominance of English keywords used to collect the dataset in our study. 

This work identifies important differences in the information space of polarized and partisan users. Better understanding how information is received, and how it propagates, can help public health experts craft more effective messaging. There are several important avenues for future work, such as designing effective interventions for misinformation, assessing the relationship between partisan asymmetries and the binding dimensions of moral thinking such as loyalty, authority and purity, and studying the temporal dynamics of these echo chambers.

\subsection {Acknowledgments}
\bibliography{main}

\end{document}